\documentstyle[12pt,epsf]{article}

\newcommand{\bmat}{\left(\begin{array}}
\newcommand{\emat}{\end{array}\right)}

\def\yzero{\smash{\hbox{$y\kern-4pt\raise1pt\hbox{${}^\circ$}$}}}

\def\-{\hphantom{-}}

\def\s2{\frac{1}{\sqrt2}}

\def\eq{\begin{equation}}
\def\eeq{\end{equation}}
\def\eqa{\begin{eqnarray}}
\def\eeqa{\end{eqnarray}}

\def\IF{\relax{\rm I\kern-.18em F}}
\def\II{\relax{\rm I\kern-.18em I}}
\def\IP{\relax{\rm I\kern-.18em P}}
\def\IC{\relax\hbox{\kern.25em$\inbar\kern-.3em{\rm C}$}}
\def\IR{\relax{\rm I\kern-.18em R}}

\def\Dsl{\,\raise.15ex\hbox{/}\mkern-13.5mu D} 
\def\IZ{Z\kern-.4em Z}

\def\pref#1{(\ref{#1})}

\def\ol#1{\overline{#1}}

\def\lrderiv#1{{\leftrightarrow 
\atop { }} {\kern-0.8em \partial_#1} \,}
\def\lrdel{{\leftrightarrow 
\atop { }} {\kern-0.8em \nabla} \,}
\def\lrdelsl{{\leftrightarrow 
\atop { }} {\kern-0.8em \delsl} \,}
\def\sss{\scriptscriptstyle}
\def\ssd{{\sss D}}
\def\ssf{{\sss F}}
\def\ssl{{\sss L}}
\def\ssp{{\sss P}}
\def\ssr{{\sss R}}
\def\sst{{\sss T}}
\def\ssv{{\sss V}}
\def\ssw{{\sss W}}
\def\ssy{{\sss Y}}
\def\ssz{{\sss Z}}

\def\fbar{\ol{f}}

\def\gbar{\ol{g}}

\def\sw{s_w}
\def\cw{c_w}

\def\mz{M_\ssz}

\def\Sca{{\cal A}}
\def\Scl{{\cal L}}
\def\roughly#1{\mathrel{\raise.3ex
\hbox{$#1$\kern-.75em\lower1ex\hbox{$\sim$}}}}
\def\lsim{\roughly<}
\def\gsim{\roughly>}

\topmargin
0.5cm
\textwidth
17.0cm
\textheight
21.5cm
\oddsidemargin
0.2cm
\evensidemargin
0.2cm

\begin{document}

\makeatletter
\@addtoreset{equation}{section}
\makeatother
\renewcommand{\theequation}{\thesection.\arabic{equation}}
\pagestyle{empty}
\vspace*{-1.0in}
\rightline{AMES-HET-00-08}
\rightline{McGill-00/20}
\rightline{UdeM-GPP-TH-00-76}
\vspace{0.5cm}
\begin{center}
 \LARGE{Supersymmetric Large Extra Dimensions\\
 Are Small and/or Numerous\\[5mm] } 
 \large{D. Atwood${}^a$, C.P. Burgess${}^b$, E. Filotas${}^b$, F. Leblond${}^b$,\\
 D. London${}^c$ and I. Maksymyk${}^d$}
 \\[4mm]
 \small{
 ${}^a$ Department of Physics and Astronomy, Iowa State University,
 Ames, Iowa 50011, USA.
 \\[1mm]
 ${}^b$ Physics Department, McGill University,\\[-0.3em]
 3600 University St., Montr\'eal, Qu\'ebec, Canada, H3A 2T8.
 \\[1mm]
 ${}^c$ Laboratoire Ren\'e J.-A. L\'evesque,
 Universit\'e de Montr\'eal, \\[-0.3em]
 C.P. 6128, succ. centre-ville, Montr\'eal, Qu\'ebec, Canada, H3C 3J7.
 \\[1mm]
 ${}^d$ Department of Physics, University of Maryland, College
 Park, Maryland 20742, USA.\\[5mm]}
 \small{\bf Abstract} \\[5mm]
\end{center}
\begin{center}
\begin{minipage}[h]{14.0cm}
    {\small Recently, a scenario has been proposed in which the
    gravitational scale could be as low as the TeV scale, and extra
    dimensions could be large and detectable at the electroweak scale.
    Although supersymmetry is not a requirement of this scenario, it
    is nevertheless true that its best-motivated realizations arise in
    supersymmetric theories (like M theory). We argue here that
    supersymmetry can have robust, and in some instances fatal,
    implications for the expected experimental signature for TeV-scale
    gravity. The signature of the supersymmetric version of the
    scenario differs most dramatically from what has been considered
    in the literature because mass splittings within the gravity
    supermultiplet in these models are extremely small, implying in
    particular the existence of a very light spin-one superpartner for
    the graviton. We compute the implications of this graviphoton, and
    show that it can acquire dimension-four couplings to ordinary
    matter which can strongly conflict with supernova bounds.}
\end{minipage}
\end{center}
\newpage
\setcounter{page}{1}
\pagestyle{plain}
\renewcommand{\thefootnote}{\arabic{footnote}}
\setcounter{footnote}{0}

\section{Introduction}

The recent transformation of the caterpillar of string theory into the
butterfly of M-theory \cite{MTheory} has drawn considerable attention
to the possibility that all known nongravitational particles might be
trapped on a 4-dimensional surface, or brane, within a
higher-dimensional `bulk' spacetime \cite{ADD,RS}. Most interestingly,
the sizes of the extra dimensions could be quite large in this
scenario, and the scale, $M_s$, describing gravitational interactions
could be as low as several TeV. Because the fundamental scale might
be so low, this is the first string `revolution' to have reached
experimental circles, stimulating considerable study of the potential
experimental signatures of the extra-dimensional graviton
\cite{realgraviton,virtualgraviton}, which is present in all such
models.

Our purpose here in this article is to explore the consequences of
another particle --- a {\it bosonic} superpartner of the graviton ---
whose existence in this scenario is almost as robust as that of the
graviton itself. All that is required is the presence of
supersymmetry. Furthermore, this particle is extremely light
(potentially as low as $m \lsim 1$ eV) even if all supersymmetries are
spontaneously broken. Although, strictly speaking, supersymmetry is
not a necessary requirement for brane models, it is a feature of
M-Theory, as well as being present in an overwhelming majority of the
explicit field-theory realizations of the brane-world scenario which
have been proposed.

Supersymmetry enters the phenomenology of brane models in an unusual
way. This is because such models deal with the heirarchy problem
differently than had been done previously. In particular, there is no
requirement that the mass splitting between particles and their
superpartners be much smaller than the gravitational scale.
Furthermore, branes themselves tend to break supersymmetries, and so
could naturally produce a nonsupersymmetric low-energy spectrum at
scales below a TeV. It is tempting to conclude from these
considerations that supersymmetry should therefore be irrelevant to
--- or have very model-dependent implications for --- experiments.

The robust signal which we explore here is based on the observation
that even if the supersymmetry breaks strongly on the brane, at scale
$M_s$, the mass splittings which this implies within the graviton
supermultiplet are {\it much} smaller, of order $M^2_s/M_p$, where
$M_p \sim 10^{19}$ GeV is the usual four-dimensional Planck mass.
Since the graviton is massless, this sets an upper bound to the masses
of its superpartners. This upper bound can be extremely low, being
$\lsim 1$ eV if $M_s \sim 1$ TeV.

Furthermore, because the graviton lives in the bulk space, its
supermultiplet is a supermultiplet for a higher-dimensional
supersymmetry. From the four-dimensional perspective, this means that
the graviton supermultiplet is necessarily a representation of {\it
 extended} supersymmetry, having at least two supersymmetry
generators. The smallest such multiplet involves two spin-3/2
gravitini, $\psi^a_\mu$, and one spin-1 boson, $V_\mu$, in addition to
the graviton.\footnote{We emphasize that this vector is not the
 extra-dimensional components of the metric tensor, which is present
 whether or not the theory is supersymmetric.} We are therefore led
on quite general grounds to the existence of a very light bulk-space
spin-one boson --- which we call the graviphoton, following the early
extended-supersymmetry literature --- in addition to the graviton and
the (two or more) gravitini. Better yet, the couplings of this spin-1
particle can be universal and of gravitational strength, since they
are related to those of the graviton by supersymmetry.

In this paper we focus on the properties of the graviphoton, and
explore its experimental implications. We identify its dominant
low-energy couplings and show how these change the predictions of
graviton exchange. Depending on the values taken by these couplings,
graviphoton exchange can alter the expected signal of large extra
dimensions quite dramatically. In particular, low-energy
Standard-Model gauge symmetries permit the graviphoton to have
dimension-four interactions with Standard-Model particles once
supersymmetry breaks, and so these couplings can dominate those of the
graviton at low energies. If such couplings are present, we find that
the rate they predict for graviphoton emission can have catastrophic
effects for supernovae, implying bounds for the underlying gravity
scale, $M_s$, which can be as large as $10^{10}$ GeV for $n=2$ extra
dimensions. TeV-scale strings can be ruled out in this case for $n <
5$.

We present our results as follows. First, in Sec.~2, we outline the
properties of the graviton supermultiplet and its couplings to
ordinary particles. We pay particular attention to the
lowest-dimension couplings it may have, and how large these might be
expected to be from the microscopic point of view. We identify two
kinds of couplings to fermions, which differ by whether or not the
corresponding interactions are derivatively coupled. The nonderivative
couplings are the ones which are potentially dangerous for supernovae,
and we estimate their size in this section. Next, in Sec.~3, we
compute the graviphoton production rate in fermion-antifermion
collisions. If the lowest-dimension couplings are suppressed by
derivatives, then we find production rates which are of order $\sigma
\sim E^n/M_s^{n+2}$, comparable to those computed elsewhere for
gravitons \cite{realgraviton}. If they are not so suppressed, then the
production rates are instead $\sigma \sim E^{n-2}/M_s^n$, and so can
be much higher at low energies. Sec.~4 then calculates the
implications of these production cross sections for supernova bounds,
with the result that non-derivative couplings are very strongly
constrained.  Sec.~5 describes the effects of virtual graviphoton
exchange on fermion scattering, and examines potential signals in
accelerator experiments.  Although the graviphoton is easily
distinguished from the graviton \cite{virtualgraviton} in these
processes, both cannot be disentangled from other states having mass
$M_s$ (such as string oscillator modes \cite{OpPollution}),
complicating the statement of what experiments should see. Finally,
our conclusions are summarized in Sec.~6.

\section{Graviphoton Properties}

We first describe the mass and couplings of the graviphoton, $V_\mu$,
focussing on those features which are model independent. We do so
first from a low-energy four-dimensional perspective, and then repeat
the discussion from a higher-dimensional point of view. As we will
see, although one can make fairly general statements about the size of
the graviphoton couplings to ordinary matter, specific details are
necessarily model dependent. (This is in contrast to graviton
couplings to matter, which are determined by the general covariance of
the theory.) Furthermore there are, unfortunately, no realistic brane
models, so one cannot even present a sample calculation of the
graviphoton couplings. Thus, at present, the discussion of these
couplings remains essentially at the level of dimensional analysis.

\subsection{Graviphoton Mass}

The most robust prediction possible is for the size of the graviphoton
mass, since this follows purely on the grounds of symmetry. (The mass
of interest in the present instance is the extra-dimensional mass,
$m_\ssv$, which is common to all Kaluza-Klein modes.) The extended
supersymmetry of the bulk space (or gravitational sector) implies two
key properties for the graviphoton. First, since it is related to the
massless graviton by supersymmetry, this graviphoton mode remains
massless so long as the four-dimensional extended supersymmetry
remains unbroken. Second, its coupling to any supersymmetry-breaking
sector is of gravitational strength. It follows that if supersymmetry
is broken by an expectation value of order $M_s^2$, then $m_\ssv \lsim
M_s^2/M_p$. For $M_s \sim 1$ TeV we have $m_\ssv \lsim 10^{-3}$ eV.
 
Because the graviphoton lives in the bulk, from the four-dimensional
viewpoint it breaks up into a tower of Kaluza-Klein (KK) states.  Each
of these states has a mass of order
\eq
\label{massform}
M_l^2 = m_\ssv^2 + {k_l \over r^2}, 
\eeq
where $r$ is a measure of the linear size of the extra dimensions and
$k_l$ is a constant depending on the extra-dimensional geometry and on
the quantum numbers, $l$, of the mode in question. Since $r$ is
related to the brane scale, $M_s$, and the Planck mass, $M_p$, by
\eq
\label{rsize}
M_p \sim (M_s \, r)^{n/2} \; M_s,
\eeq
for $n$ extra dimensions, we typically have the heirarchy $m_\ssv \ll
1/r \ll M_s$.

\subsection{Effective Trilinear Couplings}

At a microscopic level, given an explicit brane model, the graviphoton
couplings to particles may be read directly off from the model's
action. These couplings may then be evolved down to the effective
four-dimensional effective theory describing the particles we know by
integrating out all of the model's heavy or invisible degrees of
freedom. Given that the brane scale, $M_s$, is a TeV or more, it is
only the effective couplings of this low-energy theory which are
probed in experiments.

When discussing graviphoton couplings to ordinary matter it is
therefore useful to separate the discussion into two parts. First, we
may identify those effective interactions which have lowest dimension
and which are consistent with all of the known low-energy symmetries.
These will govern the phenomenological properties of the graviphoton.
We identify these couplings in the present section. Next, we may ask
how big these couplings are predicted to be in realistic brane models.
This is the topic of the next subsections. The absence of detailed
realistic constructions \cite{fernando} makes the second step more
difficult to perform, and we therefore limit ourselves in these
sections to what may be said on fairly general grounds.

In the effective four-dimensional theory the graviphoton may be
regarded as a tower of $U(1)$ gauge bosons, $V_l^\mu$ (with
$l=1,2,\dots$ labelling the various KK modes), whose gauge symmetries
commute with the electroweak gauge group. At scales below the
electroweak symmetry-breaking scale, the effective theory need not
linearly realize $SU_\ssl(2) \times U_\ssy(1)$, and so the
lowest-dimension couplings arise at dimension 4:\footnote{Our
  conventions are: $\eta_{\mu\nu} = \hbox{diag}(-++\cdots +)$ and
  $\ol\psi = i \psi^\dagger \gamma_0$.}
\eq
\label{dimfour}
\Scl_{\rm dim\, 4} = i \ol\psi \gamma_{\mu} (\hat g^l + \hat g^l_5
\gamma_5) \psi' \; V_l^\mu - {\hat\epsilon_{la} \over 2} \;
F^a_{\mu\nu} \, V_l^{\mu\nu} - \hat\lambda_l \; W_\mu^* \, W_\nu \;
\partial^\mu V_l^\nu ,
\eeq
where $\psi, \psi'$ denote the various fermions of the effective
theory, and $F^a_{\mu\nu}$ is the field strength for any of the
low-energy spin-one particles, which are labelled by the index `$a$'
({\it e.g.:} $a$ could represent the photon and the $Z$, if these are
all in the effective theory). Here and in what follows a caret over a
coupling constant is meant to emphasize that it is the
four-dimensional coupling to a particular KK mode, as opposed to the
higher-dimensional coupling between the respective fields.

Note that the coupling of two graviphotons to an ordinary gauge boson
is suppressed on general grounds by Yang's theorem, which precludes
low-energy interactions of the form $XVV$, where $X = Z, W$ or
$\gamma$. The lowest-dimension trilinear coupling which is possible is
that shown in Eq.~(\ref{dimfour}), where $\hat\lambda_l \lsim e^2
v^2/(M_s M_p)$ is an upper limit for the order of magnitude expected
for the relevant effective coupling of any one KK mode. Here $v = 246$
GeV is the order parameter for electroweak gauge-symmetry breaking,
and appears because of the necessity to break this symmetry in order
to generate the interaction in Eq.~(\ref{dimfour}).

The quantities $\hat g^l$ and $\hat g^l_5$ describe the quantum
numbers of the low-energy fermions under the $U(1)$ groups gauged by
the graviphotons, $V_l^\mu$, and would all vanish if all such fermions
were neutral under these $U(1)$'s. Even if this were so, they would be
regenerated if the couplings $\hat\epsilon_{la}$ were nonzero, since
these induce a mixing amongst the spin-one particles which can induce
a coupling of the $V_l^\mu$'s to ordinary fermions. For instance, for
energies for which we can neglect the masses of the graviphoton modes,
diagonalization of graviphoton-photon mixing produces a coupling of
the form
\eq
\label{inducedcharge}
\Scl_{\rm ind} = - \; {\hat\epsilon_{l\gamma} \over 2} \; V_l^\mu
J_\mu^{\rm em},
\eeq
where $J^{\rm em}_\mu$ denotes the usual electromagnetic current,
resulting in contributions to the $V$-fermion couplings of order
$\delta \hat g^l_5 = 0$, $\delta \hat g^l = - \frac{e}{2} \,
\hat\epsilon_{l\gamma} \, Q$, with $Q$ denoting the fermion electric
charge (normalized such that $Q=1$ for the proton).

Since dimension-four couplings are dimensionless in four spacetime
dimensions, it is tempting to hope that the couplings $\hat g, \hat
g_5$ and $\hat\epsilon$ might be unsuppressed by powers of $1/M_p$.
Although logically possible, this hope is unlikely to be fulfilled in
real brane models, as we shall see once the microscopic origin of
these couplings is traced to its higher-dimensional roots. In
subsequent sections we instead find these couplings naturally have the
size $(M_s \, r)^{-n/2} \sim M_s/M_p$. Although these are extremely
small couplings they convert, as usual, to powers of $M_s$ once summed
over the contributions of {\it all} KK modes.

For energies above the electroweak symmetry-breaking scale, but below
the threshold for other particles --- such as for superpartners if
there is a low-energy four-dimensional supersymmetry --- interactions
like those of Eq.~(\ref{dimfour}) are also possible provided that they
also linearly realize the electroweak gauge group. The possible
dimension-four interactions then become:
\eq
\label{dimfourginv}
\Scl_{\rm dim\, 4} = i \ol\psi \gamma_{\mu} (\hat{\tilde{g}}^l +
\hat{\tilde{g}}^l_5 \gamma_5) \psi' \; V_l^\mu -
{\hat{\tilde{\epsilon}}_{l} \over 2} \; B_{\mu\nu} \, V_l^{\mu\nu} ,
\eeq
where all fermions in the same electroweak representation must carry
the same graviphoton charge, and $B_{\mu\nu}$ is the field strength
for the $U_\ssy(1)$ factor of the electroweak gauge group.
Eq.~(\ref{dimfourginv}) implies Eq.~(\ref{dimfour}) with relations
amongst the effective couplings, such as $\hat\epsilon_{l\gamma} =
\hat{\tilde{\epsilon}}_l \, \cw$ and $\hat\epsilon_{l\ssz} = -
\hat{\tilde{\epsilon}}_l \, \sw$, where $\sw = \sin\theta_\ssw$ (or
$\cw = \cos \theta_\ssw$) is the sine (or cosine) of the usual weak
mixing angle.

The graviphoton coupling to gauge bosons, $\hat\lambda_l$, does not
arise at all at dimension four, but does appear once dimension-six
electroweak-invariant effective operators are considered. This leads
to the previously-quoted estimate of the coefficient $\hat\lambda_l
\sim e^2 v^2/(M_s M_p)$, where $v = 246$ GeV is the usual Higgs
expectation value.

More interactions are possible at dimension five, of which we consider
here only the fermion-graviphoton couplings as being of most practical
interest. These can have the electric/magnetic moment form:
\eq
\label{dimfive}
\Scl_{\rm dim\, 5} = i\ol\psi \gamma_{\mu\nu} (\hat a_l + i\hat b_l
\gamma_5) \psi' \; \partial^\mu V_l^\nu + \hbox{other terms}.
\eeq

Unless this interaction is forbidden by a low-energy symmetry, it can
be expected to be present with at most gravitational strength, $\hat
a_l, \hat b_l \lsim 1/M_p$, although it is typically smaller than this
upper limit.  There are at least two low-energy symmetries which can
act to dramatically reduce the size of these couplings. First, to the
extent that graviphoton interactions preserve CP (as might be expected
if they are related to graviton couplings by supersymmetry), the
effective couplings $\hat b_l$ must vanish.

Second, since the Standard-Model fermion quantum numbers forbid
writing a similar dimension-five term which linearly realizes the
electroweak gauge symmetry, an electroweak-invariant version of
Eq.~(\ref{dimfive}) first arises at dimension six. If only Standard
Model fermions (or others for which Eq.~(\ref{dimfive}) is also
forbidden) are present in the low-energy theory, then the effective
couplings $a_l$ and $b_l$ should be at most $a_l, b_l \lsim v/(M_s
M_p)$. Of course further suppressions, such as by small mixing angles
or loop factors, are also not ruled out.
 
As we shall see, there {\it are} microscopic theories for which some
fermions can acquire couplings $\hat a_l \sim O(1/M_p)$.  This can
happen if the low-energy theory may contain nonstandard fermions ---
such as gauginos or mirror fermions, for instance --- whose quantum
numbers permit forming the interaction of Eq.~(\ref{dimfive}) without
violating electroweak invariance. Since the effective couplings which
these imply for ordinary fermions are typically suppressed by powers
of small mixing angles or loop factors, we regard the estimate $\hat
a_l, \hat b_l \lsim v/(M_s M_p)$ as a fairly optimistic upper limit.

\subsection{Microscopic Couplings}

We now consider how the above effective couplings might arise from
more microscopic, higher-dimensional physics. In the absence of
detailed realistic brane models \cite{fernando} our discussion is
necessarily tentative, and we focus only on drawing conclusions which
follow from reasonably general arguments. For simplicity we also
restrict ourselves here to the more traditional scenario, with
TeV-scale gravity associated with very large extra dimensions
\cite{ADD}, as opposed to the alternative scenario \cite{RS} in which
TeV-scale gravity arises with gravity localized near our brane.

The size expected for the low-energy effective couplings described
above largely depends on how the model in question answers the
following two questions:
\begin{enumerate}
\item {\it Which of the observed low-energy particles are confined to
    the brane?} The possibility of lowering the gravity scale below
  $M_p$ ultimately arises because spin-one particles are trapped on
  our brane while gravitational forces are not. This allows long-range
  gauge forces, like electromagnetism, to `see' fewer large dimensions
  than does gravity, and so changes the connection between Newton's
  constant (or $M_p$) and the fine-structure constant, $\alpha$, on
  one hand, and the underlying coupling constants and gravity scale,
  $M_s$, on the other. We therefore require all light spin-one
  particles to be confined to our brane.
 
  It is then natural, and is usually taken to be true in the
  literature, to assume that all light fermions should also be
  confined to the brane (see, however, Ref.~\cite{bulknus}). Although
  this is the simplest possibility, it is not logically required, and
  so in what follows we contrast the usual assumption with what would
  heppen if some light fermions were to live in the bulk.
\item {\it How many supersymmetries survive below $M_s$ on our brane?}
  Since our starting point in this paper assumes the underlying theory
  is supersymmetric we must ask how many supersymmetries survive
  amongst observable particles in the low-energy theory. Although the
  splitting of the gravity multiplet in the bulk space is necessarily
  suppressed by gravitational strength interactions, there are two
  main alternatives for the splitting of supermultiplets on the brane:
\begin{enumerate}
\item {\it Supersymmetry is completely broken on our brane at scale
    $M_s$}, and so observable brane states do not have superpartners
  in the low-energy theory. In this instance the low-energy brane
  spectrum could be just the observed Standard Model fields and no
  more, although other exotic light states are logically possible.
\item {\it At least one supersymmetry survives on the brane below the
    scale $M_s$}, and so the low-energy effective theory contains
  superpartners for all of the observed particles. In this case the
  low-energy brane world might be expected to resemble the
  supersymmetric Standard Model, although probably with a specific
  form for the soft supersymmetry breaking interactions. Again the
  existence of exotic low-energy supermultiplets is not ruled out in
  this scenario.
\end{enumerate}
\end{enumerate}

\subsubsection{Couplings to Brane Particles}

First imagine that all of the relevant low-energy particles are
confined to our brane (except, of course, for the graviton
supermultiplet). To be concrete, suppose that we denote the spacetime
coordinates by $\{x^\mu, y^m\}$, where the $x^\mu$ label the
dimensions parallel to our brane, and $y^m$ label the transverse
dimensions. We further choose the position of our own brane to be
given by the condition $y^m = 0$. In this case then we must restrict
all graviphoton couplings to have the form:
\eq
\label{braneform}
\Scl_{4+n} = -\frac{1}{4} \; V_{\mu\nu}\, V^{\mu\nu} +
\Scl_{4}[\psi(x),V(x,y);M_s] \; \delta^n(y),
\eeq
where $\psi(x)$ collectively denotes all brane fields, while
$V^\mu(x,y) = \sum_l V^\mu_l(x) \, u_l(y)$ denotes the
extra-dimensional graviphoton field, having field strength $V_{\mu\nu}
= \partial_\mu V_\nu - \partial_\nu V_\mu$. The functions $u_l(y)$
here denote an appropriate complete set of Kaluza-Klein mode functions
for the extra dimensions, most of whose details play no role in what
follows \cite{frederic}. They are assumed to be normalized to preserve
the canonical normalization of the KK modes, $V_l^\mu$, {\it i.e.:}
\eq
\label{normalzn}
\int d^n y \; u_k(y) \, u_l(y) = \delta_{kl},
\eeq
where $d^n y$ denotes the appropriate covariant measure. For the
present purposes it suffices to notice that this normalization implies
the mode functions, $u_l(y)$, scale like $r^{-n/2}$ if $r$ denotes the
linear size of the extra dimensions.

The properties assumed for $\Scl_4$ are these: ($i$) it is assumed to
contain only $M_s$ as its basic dimensional scale; ($ii$) it is
assumed to linearly realize the electroweak gauge group, $SU_\ssl(2)
\times U_\ssy(1)$; ($iii$) if the low-energy brane theory should also
contain superpartners for the Standard Model fields, then $\Scl_4$ is
assumed to be supersymmetric. With these assumptions the expected
sizes for the low-energy effective couplings can be made explicit.

Suppose first that only Standard Model fields are included. In this
case no symmetries forbid the appearance of the dimension-four
couplings of Eq.~\pref{dimfourginv}. Whether they appear or not at
tree level therefore depends explicitly on whether or not Standard
Model particles carry nonzero charges for the graviphoton's $U(1)$
gauge symmetry. If these are nonzero, then dimensional analysis
determines the size of the effective four-dimensional couplings as
follows. Since these couplings are linear in $V^\mu$ they must be
proportional to one factor of the mode functions $u_k(y)$, which must
be evaluated at $y=0$ because of the delta function in
Eq.~\pref{braneform}. It follows that the couplings scale with the
radius of the extra dimensions in the same way as do the $u_k$'s:
$\hat g_l \sim g \, r^{-n/2}$, where $g$ denotes the graviphoton
coupling constant in the higher-dimensional theory. Since $M_s$
supplies all of the missing dimensions by assumption, dimensional
analysis implies $g \propto M_s^{-n/2}$ and so we see that $\hat g_l,
\hat g_{5l}$ (and $\hat\epsilon$) are all of order $(M_s r)^{-n/2}
\sim M_s/M_p$, as advertised earlier. If the extra dimensions are
torii then $u_k(y=0)$ is also independent of the mode index, $k$,
implying trilinear couplings of equal strength to all of the KK modes.

It should be remarked that even if the graviphoton charges for the
observed fermions should vanish for Standard Model fermions in the
underlying higher-dimensional theory, they will be generated through
loops unless this is precluded by another symmetry of the brane model
of interest. A concrete way in which loop-generated charges could
arise concretely, for instance, is through the generation of the
off-diagonal vacuum polarization $B_{\mu\nu} \; V^{\mu\nu}$ by loops
of any brane states which do carry nonzero charges for both
graviphotons and ordinary $U_\ssy(1)$ hypercharge. Notice that the
particles circulating in this loop can have large masses, of order
$M_s$ or larger, because their contributions to the kinetic terms need
not decouple since these have dimension four in the four-dimensional
theory. In this case one might expect $\hat e g \sim \hat \epsilon$ to
be suppressed by loop factors, although {\it not} by powers of the
mass of the particle in the loop.

Similar considerations apply to the higher-dimensional interactions.
In this case direct magnetic/electric moment graviphoton couplings to
fermions of the form of Eq.~\pref{dimfive} are precluded until
dimension six, since a power of the Higgs field is required by
electroweak gauge invariance. One therefore expects couplings of order
$(v/M_s^2) \; (M_s r)^{-n/2} \sim v /(M_s M_p)$, as advertised
earlier. Additional suppressions by mixing angles or loops may also be
present, depending on how the couplings to Standard-Model fermions
arise.

The suppression of the magnetic-moment type interactions by powers of
$v$ need not arise if nonstandard fermions are also present in the
spectrum on the brane. For instance if the weak-scale theory contains
mirror fermions in addition to those of the Standard Model, then these
fermions and their mirror pairs can form magnetic moment graviphoton
couplings without gauge invariance requiring the presence of a Higgs
{\it v.e.v.}, $v$.  The same is true for sfermions, such as gauginos,
which lie in real representations of the electroweak gauge group.
Fermions such as these are allowed interactions as in
Eq.~\pref{dimfive} with couplings of order $(M_s r)^{-n/2} \sim
M_s/M_p$, without the additional factor of $v/M_s$. Other ordinary
fermions can then acquire both dimension five and dimension four
graviphoton couplings through mixings or loops involving these
particles.

\subsubsection{Couplings to Bulk Particles}

It is not ruled out that some of the more weakly-interacting
low-energy particles --- such as right-handed neutrinos, for instance
\cite{bulknus} --- might arise microscopically as particles which can
propagate off the brane into the bulk. In this section we estimate the
sizes to be expected for the effective couplings if some fermions are
bulk-space fields.

For bulk space fields the $r$-dependence of the effective couplings
differs from that predicted by Eq.~\pref{braneform} in two ways.
First, these interactions involve an integration over the extra
dimensions due to the absence of the delta function $\delta^n (y)$.
Second, the bulk-space fields are themselves KK towers of modes,
$\psi(x,y) = \sum_l \psi_l(x) \eta_l(y)$, with canonical normalization
requiring $\eta_l \propto r^{-n/2}$, just as for the graviphoton. (It
is the presence of these KK modes, split only by masses of order
$1/r$, which permits this possibility only for sufficiently
weakly-coupled fermions.)

The couplings of such modes are also more restricted since they are
constrained by all of the spacetime symmetries and supersymmetries of
the bulk space itself. Since the bulk space enjoys extended
supersymmetries from the four-dimensional point of view, the effective
couplings of such states to the graviphoton are easily identified
using the known properties of extended supersymmetric models in four
dimensions.

In this case the tree-level effective couplings may be obtained from
inspection of the $N=2$ supergravity couplings in four spacetime
dimensions \cite{fourdsusy}. We find in this way vanishing
dimension-four couplings, $g = g_5 = \epsilon = 0$, and universal
magnetic moment couplings:
\eq
\label{vcoupl2}
\Scl_v = i\kappa_v \ol\psi \gamma^{\mu\nu} \psi' \; \partial_\mu V_\nu , 
\eeq
with $\kappa_v = c/M_s^{1+n/2}$, which are related by supersymmetry
transformations of the universal gravitational couplings:
\eq
\label{hcoupl}
\Scl_h = 2i \kappa \; h^{\mu\nu} \ol\psi \gamma_\mu
\lrderiv{\nu} \psi , 
\eeq
where (for our purposes) $\kappa = 1/M_s^{1+n/2}$ can be taken to
define the scale $M_s$.

We see that the additional symmetry implies lowest-dimensional
trilinear four-dimensional graviphoton couplings which are universal
and which preserve C, P and CP. Electroweak invariance is satisfied in
this case because four-dimensional $N=2$ supermultiplets necessarily
contain only real representations of the gauge group, implying the
presence of whatever mirror particles are required to permit the
coupling of Eq.~\pref{vcoupl2}.

Notice that the couplings of Eq.~\pref{vcoupl2} {\it must} exist at
tree level for any bulk-space spin-half particles, so if any such
particles carry electroweak quantum numbers the effective
dimension-four $B_{\mu\nu}-V_{\mu\nu}$ mixing is inevitably generated
through loops.

\section{Production Rates}

We now turn to a discussion of the phenomenological consequences of
the effective couplings of Eqs.~\pref{dimfour} and \pref{dimfive}.
Direct graviphoton production is the topic of this section, while the
influence of graviphoton exchange is pursued in the Sec.~5.

Graviphoton production is very similar to graviton production, which
has been discussed extensively in the literature within the framework
of large extra dimensions \cite{realgraviton}. The observable in both
cases is missing energy produced in association with other observed
particles, such as photons. The main conclusion we draw in this
section is that graviphoton production can change the graviton signal
in an important way, for two reasons. First, the graviphoton can
couple with a strength which is less suppressed by powers of $E/M_s$
than can the graviton, and so production of graviphotons can easily
dominate the graviton signal. It is this property which plays the
central role in the next section's discussion of supernova bounds.
Second, since the graviphoton has unit spin, it contributes
differently to observables which depend on the spin of the unobserved
particle, and so the production of these two types of particles can be
distinguished from one another, in principle, should it occur at an
observable level.

\begin{figure}
\centerline{\epsfxsize=3in\epsfbox{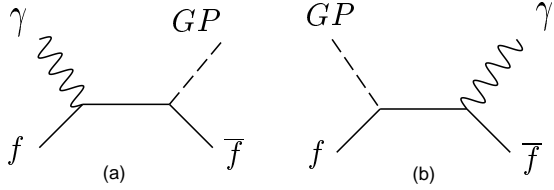}}
\begin{center}
\end{center}
\vskip-1.5truecm
\caption{\emph{The Feynman graphs contributing to photon-graviphoton
  production. The graviphoton-fermion interaction here is a sum of the
  dimension-four helicity-preserving interaction of Eq.~\pref{dimfour}
  and the dimension-five helicity-flipping interaction of
  Eq.~\pref{dimfive}.}}
\end{figure}

\subsection{Cross Section Formulae}

Graviphoton emission in fermion-antifermion collisions is mediated by
the Feynman graphs of Fig. (1), for which the graviphoton-fermion
couplings can be either the helicity-preserving dimension-four
couplings of Eq.~\pref{dimfour} or the helicity-flipping
dimension-five couplings of Eq.~\pref{dimfive}. Because of their
different helicity properties these two kinds of couplings do not
interfere with one another in the production cross section (in the
limit where fermion masses are neglected), so $d\sigma = d\sigma_{\rm
  dim \, 4} + d\sigma_{\rm dim \, 5}$.

So long as the centre-of-mass energy, $E$, of the production process
is much higher than the inverse radius of the extra dimensions, {\it
  i.e.} $E r \gg 1$, the production cross section can be computed in
either of two equivalent ways.  It may be computed in four dimensions
(by summing over the contributions of each of the many KK states) or
the calculation may be done directly in higher dimensions, with the
graviphoton represented by a single higher-dimensional field.  In
either case explicit evaluation of the graphs in Figure (1) gives the
following expressions for the (unpolarized) reaction $f\ol{f} \to
\gamma V$:
\begin{eqnarray}
\label{gravphprod}
{d\sigma_{\rm dim \, 4} \over du\, dt \, dM^2} &=& -\,
 {\alpha \, Q^2_f \, (g^2_{f} + g^2_{5f}) \over 2\, N_f \, s^2} \; 
 \left( {u\over t} + {t \over u} + {2 \, M^2 s \over u \, t }
 \right) \, P_n \, \delta(s+t+u-M^2), \\
{d\sigma_{\rm dim \, 5} \over du\, dt\, dM^2} &=& -\,
{\alpha \, Q_f^2 \, (a_f^2 + b_f^2) \over 4\, N_f \, s^2}
\; \left[ {(s^2 +M^4)\, M^2 - 4ut(u+t) \over ut} \right] \, 
P_n \, \delta(s+t+u - M^2).\nonumber \\
&& 
\end{eqnarray}
Here $g_f, g_{5f} \sim M_s^{-n/2}$ and $a_f, b_f \sim M_s^{-1-n/2}$
are the higher-dimensional coupling constants while $P_n$ denotes the
$n$-dimensional phase-space factors, $P_n = \Omega_n \, M^{n-2} /[2(2
\pi)^n]$, with $\Omega_n = 2\pi^{n/2}/\Gamma(n/2)$ representing the
volume of the $(n-1)$-sphere. $M$ represents the invariant
(four-dimensional) mass against which the photon recoils.  $N_f$ is
the number of colours carried by the initial fermion, and $s,t$ and
$u$ are the usual four-dimensional Mandelstam variables, related to
the four-momenta of the particles involved by $s = -(p_f +
p_{\ol{f}})^2$, $t = -(p_f - p_\gamma)^2$ and $u = -(p_f - p_\ssv)^2$.
This cross section was derived assuming the extra dimensions to be an
$n$-dimensional torus, although the dependence on this assumption is
weak for most values of the parameters \cite{frederic}.

\begin{figure}
\centerline{\epsfysize=3in\epsfbox{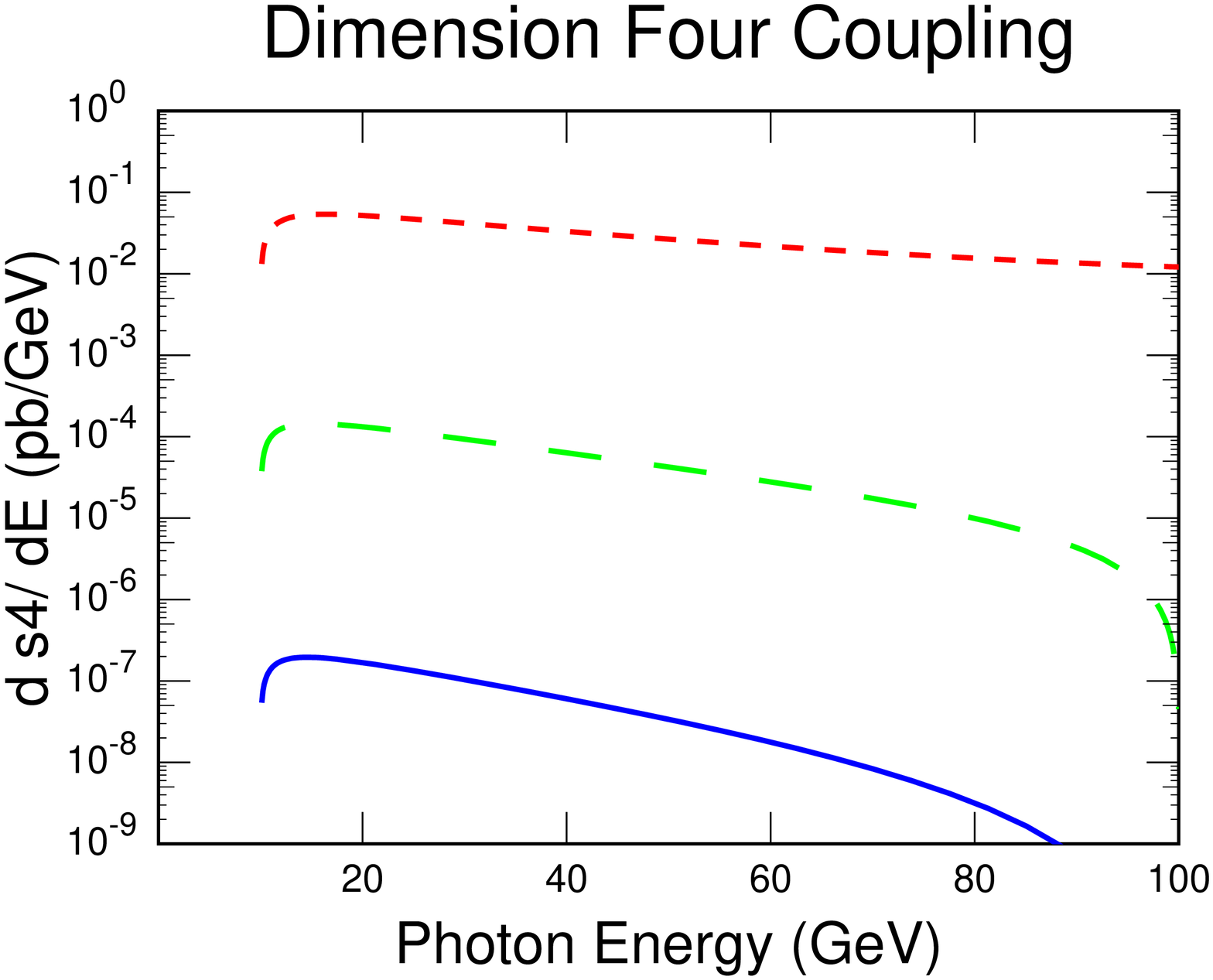}}
\begin{center}
\end{center}
\vskip-1.5truecm
\caption{\emph{The differential cross section for
  $e^+e^-\to\gamma + GP$, against photon energy, due to the
  dimension-four couplings of Eq.~\pref{dimfour}. The plot assumes an
  initial electron energy $E_e = 100$ GeV, a minimum photon transverse
  energy $\omega_\sst \ge 10$ GeV, and $M_s = 1000$ GeV. The solid
  line uses $n=6$, the long-dashed line $n=4$ and the short-dashed
  line $n=2$.}}
\end{figure}

In terms of the photon energy, $\omega$, and scattering angle,
$\theta$, relative to the incoming fermion direction (in the CM frame)
these expressions become:
\begin{eqnarray}
\label{gravphdwdc}
{d\sigma_{\rm dim \, 4} \over dx \, dc} &=&
 {\alpha \, Q^2_f \, (g^2_{f} + g^2_{5f}) \; E^{n-2} \over 
 8 \pi^{n/2}\,\Gamma(\frac{n}{2}) N_f } \; 
 \; \Bigl(1-x\Bigr)^{n/2-1} \; 
 \left({x^2(1+c^2) + 4(1-x) \over x(1-c^2)} 
 \right) , \nonumber \\
{d\sigma_{\rm dim \, 5} \over dx \, dc} &=&
{\alpha \, Q_f^2 \, (a_f^2 + b_f^2) E^n \over 2 \pi^{n/2} 
\, \Gamma(\frac{n}{2})\, N_f }
\; (1-x)^{n/2-1} \left[ {(1-x)(2-2x+x^2) + x^3 (1-c^2) 
\over x(1-c^2)} \right] ,
\end{eqnarray}
where $x = \omega/E_e$, $E_e$ is the incoming electron energy and $c
\equiv \cos\theta$.

When comparing with experiments, these expressions must be added to
those for the cross section for photon-graviton production, $f\ol{f}
\to \gamma g$, which has been calculated in Refs.~\cite{realgraviton},
whose results we reproduce here (using our conventions\footnote{For
  instance, our $\kappa^2 = 1/M_s^{n+2}$ is $(2\pi)^n/M_\ssd^{n+2}$
  with $M_\ssd$ as defined by Giudice {\it et.al.}
  \cite{realgraviton}.}) for convenience of reference:
\eq
\label{gravitonprod}
{d\sigma_{\rm g} \over dx \, dc} =
{\alpha \, Q_f^2 \, \kappa^2 \, E^n \over 16 \pi^{n/2} 
\, \Gamma(\frac{n}{2})\, N_f } \; (1-x)^{n/2-1} 
\left[ {(2-x)^2 (1-x+x^2) -3 x^2 \, c^2 (1-x) - x^4 c^4 \over x(1-c^2)} 
\right] ,
\eeq
where $\kappa = 1/M_s^{1+n/2}$.

Figures (2) through (5) plot the singly-differential cross sections,
which are obtained by integrating Eqs.~\pref{gravphdwdc}. The
collinear mass singularities associated with the neglect of the
electron mass in these expressions is excluded by the physical
requirement that the photon have a minimum transverse energy,
$\omega_\sst \ge 10$ GeV.

\begin{figure}
\centerline{\epsfysize=3in\epsfbox{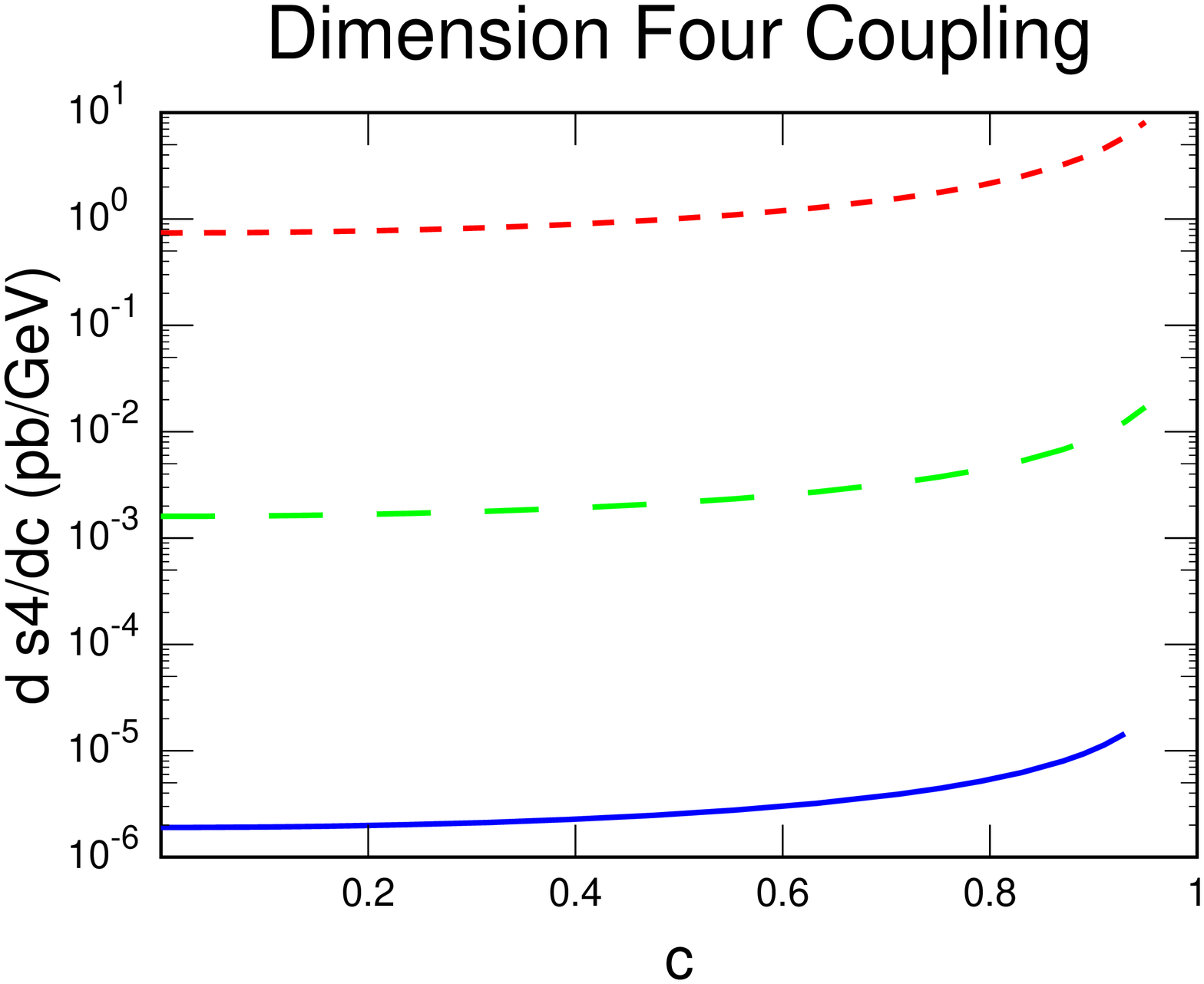}}
\begin{center}
\end{center}
\vskip-1.5truecm
\caption{\emph{The differential cross section for
  $e^+e^-\to\gamma + GP$, against photon scattering angle in the CM
  frame, arising due to the dimension-four couplings of
  Eq.~\pref{dimfour}. The plot assumes an initial electron energy $E_e
  = 100$ GeV, a minimum photon transverse energy $\omega_\sst \ge 10$
  GeV, and $M_s = 1000$ GeV. The solid line uses $n=6$, the
  long-dashed line $n=4$ and the short-dashed line $n=2$.}}
\end{figure}

The total production cross section due to the dimension-four
interactions becomes:
\eq
\label{totsig}
\sigma_{\rm dim 4} = k_n \; \left({E \over 100 \, \hbox{GeV}} \right)^{n-2}
\; \left( {g_f^2 + g_{5f}^2 \over (1 \, \hbox{TeV})^{-n}} \right),
\eeq
with 
\eq
\label{kvals}
k_n = \cases{ 3.5~{\rm pb} ~, & $n=2$ ~, \cr
 0.18~{\rm pb} ~, & $n=3$ ~, \cr
 7.5 \times 10^{-3}~{\rm pb} ~, & $n=4$ ~, \cr
 2.7 \times 10^{-4}~{\rm pb} ~, & $n=5$ ~, \cr
 8.8 \times 10^{-6}~{\rm pb} ~, & $n=6$ ~,\cr}
\eeq
while the corresponding result using the dimension-five interactions
is:
\eq
\label{totsig5}
\sigma_{\rm dim 5} = \tilde{k}_n \; \left({E \over 100 \, \hbox{GeV}} \right)^{n}
\; \left({ a_f^2 + b_{f}^2 \over (1 \, \hbox{TeV})^{-n+2}} \right),
\eeq
with 
\eq
\label{kvals5}
\tilde{k}_n = \cases{ 0.051~{\rm pb} ~, & $n=2$ ~, \cr
 2.7 \times 10^{-3}~{\rm pb} ~, & $n=3$ ~, \cr
 1.1 \times 10^{-4}~{\rm pb} ~, & $n=4$ ~, \cr
 4.2 \times 10^{-6}~{\rm pb} ~, & $n=5$ ~, \cr
 1.4 \times 10^{-7}~{\rm pb} ~, & $n=6$ ~.\cr}
\eeq

\begin{figure}
\centerline{\epsfysize=3in\epsfbox{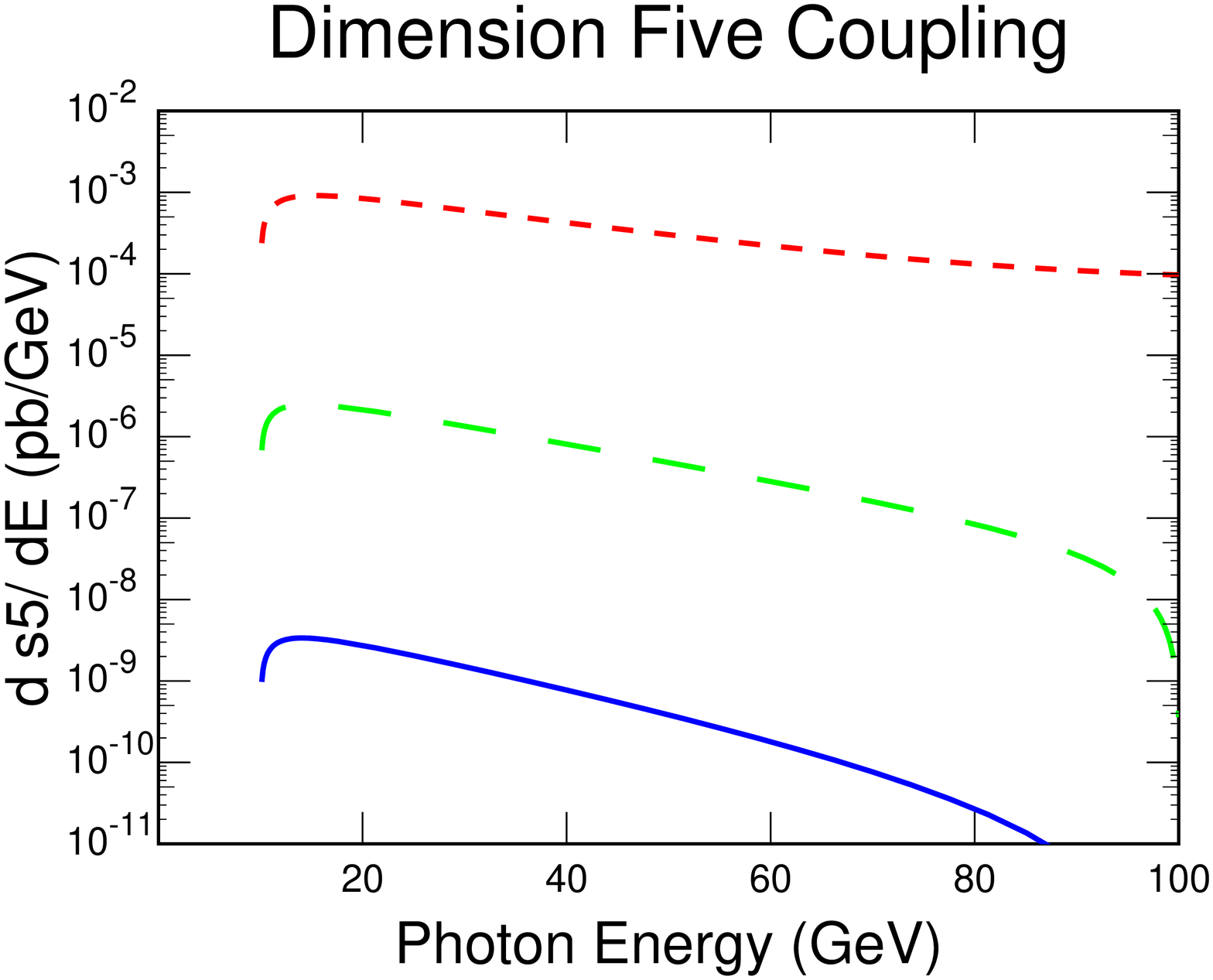}}
\begin{center}
\end{center}
\vskip-1.5truecm
\caption{\emph
{The differential cross section for
  $e^+e^-\to\gamma + GP$, against photon energy, due to the
  dimension-five couplings of Eq.~\pref{dimfive}. The plot assumes an
  initial electron energy $E_e = 100$ GeV, a minimum photon transverse
  energy $\omega_\sst \ge 10$ GeV, and $M_s = 1000$ GeV. The solid
  line uses $n=6$, the long-dashed line $n=4$ and the short-dashed
  line $n=2$.}}
\end{figure}

These results are to be compared with Standard Model background
processes, which are principally due to $\gamma-Z$ emission, with the
$Z$ subsequently decaying invisibly into neutrinos, as has been
discussed in detail for graviton production in
Refs.~\cite{realgraviton}.

\subsection{Discussion}

The property of the production cross sections which is central to the
remaining discussions is its dependence on the centre-of-mass energy.
This dependence is a simple power law -- being proportional to
$E^n/M_s^{n+2}$ for both the graviton and the magnetic-moment
graviphoton interactions. The simple power-law behaviour is typical of
low-energy effective interactions, and arises in the present case due
to the derivative nature of both of these couplings. By contrast, the
energy dependence of the dimension-four graviphoton production cross
section is $E^{n-2}/M_s^n$. This means that this coupling is
suppressed by two fewer powers of $E/M_s$ than are the others. This
observation has several consequences.

\begin{enumerate}
\item All other things being equal, at low energies the dimension-four
  graviphoton couplings must dominate, and their domination is
  stronger the lower the energy of interest. This makes them
  particularly important for supernovae bounds, for which the relevant
  processes involve energies of order 10 MeV.
\item For the same reason production rates for graviphotons should
  also dominate graviton production in accelerators, provided only
  that the dimension four couplings are not additionally suppressed
  relative to their size expected on dimensional grounds, $g \sim
  1/M_s^{n/2}$.
\item If it happens that the dimension-four couplings {\it are}
  suppressed, such as by loop or other small factors, and it is the
  magnetic moment interactions which dominate the production cross
  section, then graviphoton production may be expected to be
  comparable to graviton production. The bounds on $M_s$ from existing
  experiments, and the reach of future ones, can be expected in this
  case to be similar to those obtained using only graviton production
  \cite{realgraviton}.
\item For both graviphoton and graviton production, the cross sections
  decrease as $n$ increases. Therefore, the more large extra
  dimensions there are, the more difficult it is to observe their
  effects in experiments at energies $\sqrt{s} < M_s$.
\item In the event that we are ever lucky enough to find missing
  energy processes unexplained by the Standard Model, then the
  difference in the angular distributions between
  Eqs.~\pref{gravphdwdc} and \pref{gravitonprod} can be used to
  distinguish the contributions of each type of particle.
\end{enumerate}

\begin{figure}
\centerline{\epsfysize=3in\epsfbox[0 0 500 500]{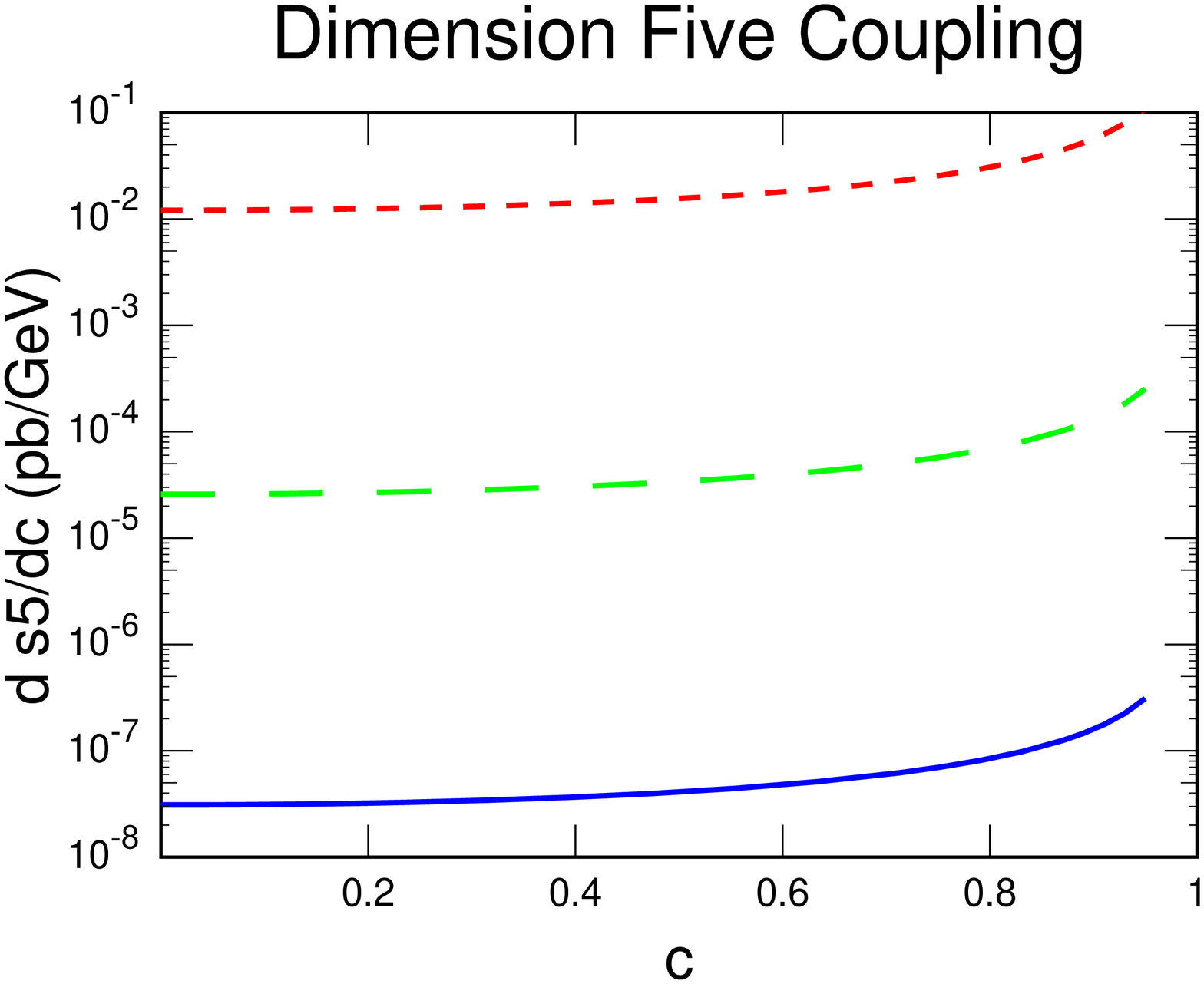}}
\begin{center}
\end{center}
\vskip-1.5truecm
\caption{\emph
{The differential cross section for
  $e^+e^-\to\gamma + GP$, against photon scattering angle in the CM
  frame, arising due to the dimension-five couplings of
  Eq.~\pref{dimfive}. The plot assumes an initial electron energy $E_e
  = 100$ GeV, a minimum photon transverse energy $\omega_\sst \ge 10$
  GeV, and $M_s = 1000$ GeV. The solid line uses $n=6$, the
  long-dashed line $n=4$ and the short-dashed line $n=2$.}}
\end{figure}

\section{Bounds from Supernovae}

In the previous section, we saw that the process $f\ol{f} \to \gamma
V$ can be used in future experiments to set limits on the gravity
scale, $M_s$. The exact bound obtained will depend on the number of
extra dimensions $n$.

Supernovae have the potential to probe much larger values of $M_s$,
because both gravitons and graviphotons can be emitted freely, thereby
increasing the rate of energy loss. Indeed, the fact that the observed
neutrino flux from the supernova SN 1987A agrees with the predictions
of stellar collapse models has been used to constrain graviton
emission. The authors of Ref.~\cite{supernovae} give the following
expressions for the rate of energy loss, ${\dot\epsilon}$, due to
graviton emission:
\eq
{\dot\epsilon} = \cases{ 
5.1 \times 10^{17} ~{\hbox{erg g$^{-1}$ s$^{-1}$}}
~T_{\rm MeV}^{5.5} \, M_{\rm TeV}^{-4} ~, & $n=2$ ~, \cr
2.8 \times 10^{12} ~{\hbox{erg g$^{-1}$ s$^{-1}$}}
~T_{\rm MeV}^{6.5} \, M_{\rm TeV}^{-5} ~, & $n=3$ ~, \cr
1.7 \times 10^{7} ~{\hbox{erg g$^{-1}$ s$^{-1}$}}
~T_{\rm MeV}^{7.5} \, M_{\rm TeV}^{-6} ~, & $n=4$ ~, \cr}
\eeq
where $T_{\rm MeV} = T/(1~{\rm MeV})$ and $M_{\rm TeV} = M_s / (1~{\rm
  TeV})$. Taking $T = 30$ MeV, and requiring that ${\dot\epsilon} <
10^{19} ~{\hbox{erg g$^{-1}$ s$^{-1}$}}$ \cite{raffelt}, one obtains a
very stringent constraint for $n=2$: $M_s \gsim 50$ TeV. For larger
values of $n$ this constraint is weaker: $M \gsim 4$ TeV (1 TeV) for
$n = 3$ ($n=4$). The conclusion is therefore that TeV-scale quantum
gravity is possible only for $n\ge 3$.

An analogous analysis can be done for graviphoton emission. An
examination of Eqs.~\pref{gravphdwdc} and \pref{gravitonprod} reveals
that the emission of a graviphoton with dimension-5 couplings is very
similar to that of a graviton, and we thus expect similar constraints.
However, we see from Eq.~\pref{gravphdwdc} that the cross section for
the production of a graviphoton with dimension-4 couplings is enhanced
relative to graviton emission by a factor $M_s^2 / s$. Since the
temperature in the core of supernovae is about $30$ MeV, this
enhancement factor is {\it enormous}, and this has profound
implications for models with graviphotons. To estimate the energy loss
for the emission of a graviphoton with dimension-4 couplings, we
naively scale the graviton results up by a factor $M_s^2 / s$. This
yields
\eq
{\dot\epsilon} = \cases{ 
5.1 \times 10^{29} ~{\hbox{erg g$^{-1}$ s$^{-1}$}}
~T_{\rm MeV}^{3.5} \, M_{\rm TeV}^{-2} & $n=2$ ~, \cr
2.8 \times 10^{24} ~{\hbox{erg g$^{-1}$ s$^{-1}$}}
~T_{\rm MeV}^{4.5} \, M_{\rm TeV}^{-3} & $n=3$ ~, \cr
1.7 \times 10^{19} ~{\hbox{erg g$^{-1}$ s$^{-1}$}}
~T_{\rm MeV}^{5.5} \, M_{\rm TeV}^{-4} & $n=4$ ~, \cr
\sim 10^{14} ~{\hbox{erg g$^{-1}$ s$^{-1}$}}
~T_{\rm MeV}^{6.5} \, M_{\rm TeV}^{-5} & $n=5$ ~, \cr
\sim 10^{9} ~{\hbox{erg g$^{-1}$ s$^{-1}$}}
~T_{\rm MeV}^{7.5} \, M_{\rm TeV}^{-6} & $n=6$ ~. \cr}
\label{gpemission}
\eeq
The expression for the energy loss for $n=5$ has been estimated by
multiplying the $n=4$ formula by a simple scale factor $\sim 10^{-5}$,
and similarly for $n=6$. The constraints on the gravity scale, $M_s$,
are now considerably more stringent:
\eq
M_s \gsim \cases{ 
10^8~{\rm TeV} ~, & $n=2$ ~, \cr
10^4~{\rm TeV} ~, & $n=3$ ~, \cr
10^2~{\rm TeV} ~, & $n=4$ ~, \cr
10~{\rm TeV} ~, & $n=5$ ~, \cr
1~{\rm TeV} ~, & $n=6$ ~. \cr}
\eeq
Thus, allowing for a factor of 10 leeway in our naive estimates, we
conclude that, for models with graviphotons with dimension-4
couplings, TeV-scale quantum gravity requires a large number of large
extra dimensions, $n\ge 5$. Conversely, if there are only a small
number of large extra dimensions, they cannot be large enough to be
detectable at future colliders.

A comparison of graviphoton and axion emission from supernovae shows
that these results are not unexpected. Since the axion has a
derivative coupling to matter, the cross section for axion emission is
proportional to $f_a^{-2}$, where $f_a$ is the axion decay constant.
Constraints on axion emission from supernovae imply that $f_a \gsim
10^{10}$ GeV \cite{axionref}. For $n=2$, the rate for graviphoton
emission is also proportional to $M_s^{-2}$ [see
Eq.~(\ref{gpemission})]. Thus, purely on dimensional grounds, we
expect that $M_s \gsim 10^{10}$ GeV, and this is indeed what is found
(to within a factor of 10).

Of course, implicit in the above analysis is the assumption that the
dimension-4 couplings are nonzero. Indeed, we have taken the couplings
$g_f, g_{5f}$ to be a number of $O(1)$ multiplied by $M_s^{-n/2}$.
One might therefore try to evade the above constraints by simply
assuming that these couplings vanish. However, this is, in general,
not possible. As discussed in Sec.~2.3.1, even if the ordinary
fermions are neutral under the graviphoton $U(1)$ gauge group,
dimension-4 couplings of the graviphoton to ordinary matter will be
``grown'' at loop level (e.g.\ due to mixing of the graviphoton and
the $B_\mu$ gauge boson). Thus, although one cannot take $g_f, g_{5f}$
to be zero, it is conceivable that they might be small, i.e.\ of
loop-level size, $10^{-2}$, multiplied by $M_s^{-n/2}$. In this case,
the rate for energy loss due to graviphoton emission in
Eq.~(\ref{gpemission}) would be reduced by a factor $\sim 10^{-4}$.
This leads to somewhat weaker constraints on the gravity scale:
\eq
M_s \gsim \cases{ 
10^6~{\rm TeV} ~, & $n=2$ ~, \cr
500~{\rm TeV} ~, & $n=3$ ~, \cr
10~{\rm TeV} ~, & $n=4$ ~, \cr
1~{\rm TeV} ~, & $n=5$ ~. \cr}
\eeq

We therefore conclude that, due to supernovae constraints on
graviphoton emission, TeV-scale quantum gravity is generically
possible only if there are at least four large extra dimensions.
Alternatively, any explicit model of underlying brane dynamics which
wishes to describe TeV-scale branes with few extra dimensions (and so
to have large signals in future colliders) must be designed to
insulate any of the graviphotons from ordinary matter at much better
than the one-loop level.

Note that these conclusions (i) assume that the dimension-4
graviphoton couplings are as small as they can be, $g \sim 10^{-2}
M_s^{-n/2}$, and (ii) allow for a factor of 10 error in our estimates
of the supernova bound. They also assume the existence of only one
light graviphoton, whereas models with many extra dimensions are
likely to have many such graviphotons due to the larger number of
bulk-space supersymmetries they involve, seen from the
four-dimensional perspective.  Our conclusions may thus be considered
to be reasonably conservative.

\section{Implications for Fermion-Antifermion Scattering}

We next turn to potential observable effects in fermion-antifermion
collisions due to graviphoton exchange, induced by the effective
couplings of Eqs.~\pref{dimfour} and \pref{dimfive}. We find that
graviphoton exchange can significantly change what would be expected
from graviton exchange.

Unfortunately, this difference between graviton and graviphoton
exhange is of less practical interest than was the difference found in
the previous section for graviphoton and graviton production. The
argument is the following. In TeV-scale quantum gravity models,
graviton exchange can significantly affect fermion-antifermion
scattering. This has generated a great deal of excitement, and such
graviton-exchange contributions have been extensively discussed in the
literature \cite{virtualgraviton}. The reason that such keen interest
was generated is that these effects were thought to be clean,
model-independent tests for the presence of large extra dimensions.
Unfortunately, this is not the case. As discussed in
Ref.~\cite{OpPollution}, graviton exchange is indistinguishable at low
energies from other, equally large, scattering contributions due to
the exchange of states having masses $M_s$, such as higher string
modes in both the brane and bulk sectors.  These other contributions
are dangerous because they can contribute to fermion scattering at the
same order in $1/M_s$ as can the graviton KK modes, and their effects
depend on the details of the microscopic spectrum of the underlying
theory. Thus, all the calculations in Ref.~\cite{virtualgraviton} are
incomplete since they neglect these other, model-dependent
contributions.

Obviously, this also applies to graviphotons. Although it is possible
to calculate the effects of graviphoton exchange on
fermion-antifermion scattering, even including the interference with
graviton exchange, such calculations will also be incomplete. Even so,
we present here the results for tree-level graviphoton exchange in
fermion-antifermion scattering. We do so partly for the sake of
completeness, and partly because these expressions are the necessary
first step towards any future understanding of how to disentangle the
various kinds of extra-dimensional effects.

\begin{figure}
\centerline{\epsfxsize=3in\epsfbox{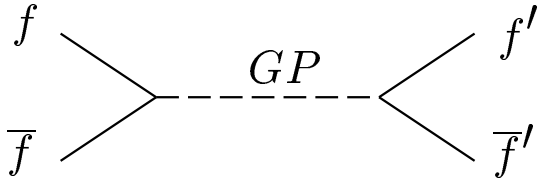}}
\medskip
\begin{center}
\end{center}
\vskip-1.5truecm
\caption{\emph
{The Feynman graphs giving the graviphoton-exchange
  contribution to fermion-antifermion scattering. As in Fig. (1), the
  graviphoton-fermion interaction here is a sum of the dimension-four
  helicity-preserving interaction of Eq.~\pref{dimfour} and the
  dimension-five helicity-flipping interaction of Eq.~\pref{dimfive}.}}
\end{figure}

\subsection{Cross Section Formulae}

The Feynman graphs which are potentially relevant for tree-level
graviphoton contributions to fermion-antifermion scattering are those
of Fig. (6). Again, the graviphoton-fermion coupling is the sum of the
helicity-preserving dimension-four terms of Eq.~\pref{dimfour} and the
helicity-flipping dimension-five terms of Eq.~\pref{dimfive}.

In the (excellent) approximation where fermion masses are neglected,
it is instructive to distinguish the cross sections for the sixteen
different combinations of helicities for the fermions in the initial
and final states. This is because only a limited number of couplings
contribute to any one helicity combination, potentially providing a
wealth of diagnostic information as to the kind of scattering which is
responsible for the observed cross section.

The results are most conveniently displayed in the following way.
Since fermion masses are neglected, the unpolarized cross section is
$\frac14$ the sum of the result for each possible choice of initial
and final helicity.  We list each of these polarized cross sections:
\eq
\label{ffmaster}
{d \sigma \over du \, dt}(f_i \fbar_j \to g_{i'} \gbar_{j'})
 = {1 \over 4 \pi \, s^2} \; \left| \Sca_{ij;i'j'}\right|^2 \; \delta(s+t+u),
\eeq
where $i,j,i',j'=L,R$ denote the helicities of the initial and final
fermions. The remainder of this section gives the relevant amplitudes,
$\Sca_{ij;i'j'}$, first for the case $g \ne f$ ({\it e.g.} for $e^+e^-
\to \mu^+\mu^-$) and then for $g = f$ (for $e^+e^- \to e^+e^-$).

\subsubsection{$e^+e^- \to \ell^+\ell^-$}

Consider first the case where the final and initial fermions are
distinguishable.  The amplitudes relevant for this case are summarized
as a function of the initial and final helicities in Table (1).

\begin{table}
\begin{center}
\begin{tabular}{|c||c|c|c|c|}
\hline
 & $\ell^-_\ssl \ell^+_\ssl$ & $\ell^-_\ssl \ell^+_\ssr$ 
 & $\ell^-_\ssr \ell^+_\ssl$ & $\ell^-_\ssr \ell^+_\ssr$ \\
\hline
\hline
 $e^-_\ssl e^+_\ssl$ & $|\Sca(\hbox{FF})|^2$ & $|\Sca(\hbox{FP})|^2$ 
 & $|\Sca(\hbox{FP})|^2$ & $|\Sca(\hbox{FF})|^2$ \\
\hline
 $e^-_\ssl e^+_\ssr$ & $|\Sca(\hbox{PF})|^2$ & $|\Sca_{\ssl\ssl}(\hbox{PP})|^2$ 
 & $|\Sca_{\ssl\ssr}(\hbox{PP})|^2$ & $|\Sca(\hbox{PF})|^2$ \\
\hline
 $e^-_\ssr e^+_\ssl$ & $|\Sca(\hbox{PF})|^2$ & $|\Sca_{\ssr\ssl}(\hbox{PP})|^2$ 
 & $|\Sca_{\ssr\ssr}(\hbox{PP})|^2$ & $|\Sca(\hbox{PF})|^2$ \\
\hline
 $e^-_\ssr e^+_\ssr$ & $|\Sca(\hbox{FF})|^2$ & $|\Sca(\hbox{FP})|^2$ 
 & $|\Sca(\hbox{FP})|^2$ & $|\Sca(\hbox{FF})|^2$\\ 
\hline
\end{tabular}
\end{center}
\caption{
\emph{The amplitude, $\Sca$, appearing in the differential cross
  section, Eq.~\pref{ffmaster}, for $e^+e^- \to \ell^+\ell^-$, for
  $\ell \ne e$, in the limit where electron and final-state lepton
  masses are neglected.  The labels `F' and `P' are meant to indicate
  helicity-flipping or helicity-preserving interactions.}}
\end{table}

The functions which appear in these tables are given explicitly by:
\begin{eqnarray}
\label{eemumuexpr1}
|\Sca(\hbox{FF})|^2 &=& \frac14 \, (a_e^2 + b_e^2) \, (a_\mu^2 + b_\mu^2) 
 \; s^2 \, (t-u)^2 \; |F|^2 \\
\label{eemumuexpr2}
|\Sca(\hbox{PF})|^2 &=& (a_\mu^2+ b_\mu^2) \, (g_{e}^2 + g_{5e}^2) \; 
 s \, t \, u \; |F|^2\\
\label{eemumuexpr3}
|\Sca(\hbox{FP})|^2 &=& (a_e^2+ b_e^2) \, (g_{\mu}^2 + g_{5\mu}^2) \; 
 s \, t \, u \; |F|^2
\end{eqnarray}
and 
\begin{eqnarray}
\label{eemumuexpr4}
|\Sca_{\ssl\ssl}(\hbox{PP})|^2 &=& u^2\, \left|
 \left[{e^2 \over s} + \left({e\over \sw\cw}\right)^2 \; 
 {h_{e\ssl}\, h_{\mu\ssl} \over s - \mz^2} \right] 
 + g_{e\ssl} \, g_{\mu\ssl} \, F 
 +2 \kappa^2 \; (4u + 3s) \; F \right|^2 \\
\label{eemumuexpr5}
|\Sca_{\ssr\ssr}(\hbox{PP})|^2 &=& u^2 \, \left|
 \left[{e^2 \over s} + \left({e\over \sw\cw}\right)^2 
 \; {h_{e\ssr}\, h_{\mu\ssr} \over s - \mz^2} \right] 
 + g_{e\ssr} \, g_{\mu\ssr} \, F 
 + 2 \kappa^2 \; (4u + 3s) \; F \right|^2 \\
\label{eemumuexpr6}
|\Sca_{\ssl\ssr}(\hbox{PP})|^2 &=& t^2\, \left| 
 \left[{e^2 \over s} + \left({e\over \sw\cw}\right)^2 \; 
 {h_{e\ssl}\, h_{\mu\ssr} \over s - \mz^2} \right] 
 + g_{e\ssl} \, g_{\mu\ssr} \, F 
 + 2 \kappa^2 \; (4u + s) \; F \right|^2 \\
\label{eemumuexpr7}
|\Sca_{\ssr\ssl}(\hbox{PP})|^2 &=& t^2 \, \left| 
 \left[{e^2 \over s} + \left({e\over \sw\cw}\right)^2 \; 
 {h_{e\ssr}\, h_{\mu\ssl} \over s - \mz^2} \right] 
 + g_{e\ssr} \, g_{\mu\ssl} \, F 
 + 2 \kappa^2 \; (4u + s)\; F \right|^2 
\end{eqnarray}
where $h_\ssl = -\frac12 + \sw^2$ and $h_\ssr = \sw^2$ denote the
couplings of left- and right-handed fermions to the $Z$ boson, while
$g_\ssl = (g + g_5)$ and $g_\ssr = (g - g_5)$ are the corresponding
dimension-four fermion couplings to the graviphoton.  All of these
expressions neglect the mass of the initial and final state fermions,
as well as the $(4+n)$-dimensional mass, $m_\ssv$, of the graviphoton.

The quantity $F$ in these expressions is the function defined by:
\eq
\label{Fdef}
F(s) = \int {d^n Q \over (2 \pi)^n} \; {1 \over Q^2 - s} ,
\eeq
and so $F \to -1/s$ as $n \to 0$ -- {\it i.e.} in the absence of extra
dimensions. Since this expression diverges in the ultraviolet, for $n
\ge 2$, we follow the practice in the literature and take it to be cut
off at the scale $Q^2 = M_s^2$, making $F$ approximately constant ($F
= \Omega_n M_s^{n-2} /[(n-2)(2\pi)^n]$ if $n>2$, up to $O(s/M_s^2)$
corrections). Because the couplings $g, g_5$ and $\kappa$ are the
higher-dimensional ones, this form for $F$ implies
\eq
\label{coupsize}
g^2 F \sim {1 \over M_s^2} \qquad \hbox{and} \qquad
a^2 F \sim \kappa^2 F \sim {1 \over M_s^4} , 
\eeq
and so the four-dimensional Planck scale, $M_p$, drops out of these
expressions in the usual way. (From the four-dimensional point of view
it does so because the factor of $1/M_p$ in the coupling cancels
factors of $M_p$ in the density of states of the KK modes.)

Notice that the necessity to cut off the integral in Eq.~\pref{Fdef}
is a symptom that the graviphoton and graviton contributions are
competing with the low-energy effects of exchanging heavier (string)
states having masses of order $M_s$. These two issues are connected
because the ultraviolet divergence we are regulating must be
absorbable into the renormalization of an effective coupling of the
low-energy theory.  (Such a renormalization is indeed possible because
the relevant terms in Eqs.~\pref{eemumuexpr1} through
\pref{eemumuexpr7} are polynomials in $s, t$ and $u$, and so
correspond to the effects of local operators.)  The couplings whose
renormalizations do the job are precisely those of the effective
interactions which are generated on integrating out the mass-$M_s$
heavy states.

\begin{figure}
\centerline{\epsfxsize=3in\epsfbox{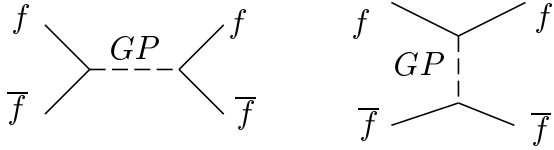}}
\medskip
\begin{center}
\end{center}
\vskip-1.5truecm
\caption{\emph
{The Feynman graphs giving the graviphoton-exchange
  contribution to fermion-antifermion scattering, when the initial and
  final fermions are indistinguishable. As for previous figures, the
  graviphoton-fermion interaction here is a sum of the dimension-four
  helicity-preserving interaction of Eq.~\pref{dimfour} and the
  dimension-five helicity-flipping interaction of Eq.~\pref{dimfive}.}}
\end{figure}

\subsubsection{$e^+e^- \to e^+e^-$}

Next consider the corresponding expressions for the case where initial
and final particles are indistinguishable, $f=g$ (and so which apply
for $e^+e^- \to e^+e^-$). The Feynman graphs relevant to this process
are given in Fig. (7), and differ from the previous case by the
inclusion of $t$-channel graphs. Neglecting the electron mass gives
the total cross section as $\frac14$ the sum over the polarized cross
sections, which we again express as in Eq.~\pref{ffmaster}. The
results for the polarized cross sections obtained from evaluating the
graphs of Fig. (7) are listed in Table (2).

\begin{table}
\begin{center}
\begin{tabular}{|c||c|c|c|c|}
\hline
 & $e^-_\ssl e^+_\ssl$ & $e^-_\ssl e^+_\ssr$ 
 & $e^-_\ssr e^+_\ssl$ & $e^-_\ssr e^+_\ssr$ \\
\hline
\hline
 $e^-_\ssl e^+_\ssl$ & $|\Sca_{\ssl\ssr}^{\ssf\ssf}(s)+\Sca_{\ssl\ssr}^{\ssp\ssp}(t)|^2$ 
 & $|\Sca_{\ssl\ssl}^{\ssf\ssp}(s)+\Sca_{\ssl\ssl}^{\ssp\ssf}(t)|^2$ 
 & $|\Sca_{\ssl\ssr}^{\ssf\ssp}(s)+\Sca_{\ssl\ssr}^{\ssf\ssp}(t)|^2$ 
 & $|\Sca_{\ssl\ssl}^{\ssf\ssf}(s)+\Sca_{\ssl\ssl}^{\ssf\ssf}(t)|^2$ \\
\hline
 $e^-_\ssl e^+_\ssr$ & $|\Sca_{\ssl\ssr}^{\ssp\ssf}(s)+\Sca_{\ssl\ssr}^{\ssp\ssf}(t)|^2$ 
 & $|\Sca_{\ssl\ssl}^{\ssp\ssp}(s)+\Sca_{\ssl\ssl}^{\ssp\ssp}(t)|^2$ 
 & $|\Sca_{\ssl\ssr}^{\ssp\ssp}(s)+\Sca_{\ssl\ssr}^{\ssf\ssf}(t)|^2$ 
 & $|\Sca_{\ssl\ssl}^{\ssp\ssf}(s)+\Sca_{\ssl\ssl}^{\ssf\ssp}(t)|^2$ \\
\hline
 $e^-_\ssr e^+_\ssl$ & $|\Sca_{\ssr\ssr}^{\ssp\ssf}(s)+\Sca_{\ssr\ssr}^{\ssf\ssp}(t)|^2$ 
 & $|\Sca_{\ssr\ssl}^{\ssp\ssp}(s)+\Sca_{\ssr\ssl}^{\ssf\ssf}(t)|^2$ 
 & $|\Sca_{\ssr\ssr}^{\ssp\ssp}(s)+\Sca_{\ssr\ssr}^{\ssp\ssp}(t)|^2$ 
 & $|\Sca_{\ssr\ssl}^{\ssp\ssf}(s)+\Sca_{\ssr\ssl}^{\ssp\ssf}(t)|^2$ \\
\hline
 $e^-_\ssr e^+_\ssr$ & $|\Sca_{\ssr\ssr}^{\ssf\ssf}(s)+\Sca_{\ssr\ssr}^{\ssf\ssf}(t)|^2$ 
 & $|\Sca_{\ssr\ssl}^{\ssf\ssp}(s)+\Sca_{\ssr\ssl}^{\ssf\ssp}(t)|^2$ 
 & $|\Sca_{\ssr\ssr}^{\ssf\ssp}(s)+\Sca_{\ssr\ssr}^{\ssp\ssf}(t)|^2$ 
 & $|\Sca_{\ssr\ssl}^{\ssf\ssf}(s)+\Sca_{\ssr\ssl}^{\ssp\ssp}(t)|^2$\\ 
\hline
\end{tabular}
\end{center}
\caption{
\emph{The amplitude, $\Sca$, appearing in the differential cross
  section, Eq.~\pref{ffmaster}, for $e^+e^- \to e^+ e^-$, in the limit
  where electron mass is neglected.  The labels `F' and `P' are meant
  to indicate helicity-flipping or helicity-preserving interactions,
  and `$s$' and `$t$' indicate the results of $s$- and $t$-channel
  graphs.}}
\end{table}

The $s$-channel amplitudes which appear in this table are given by the
following expressions (when the graviphoton and electron masses are
neglected).

\begin{eqnarray}
\label{eeeeexpr1}
\Sca_{\ssl\ssl}^{\ssf\ssf}(s) = \Bigl(\Sca_{\ssr\ssr}^{\ssf\ssf}(s)\Bigr)^* &=& 
 \frac12 \, (a_e + ib_e)^2 \; s \, (t-u) \; F(s) \nonumber\\
\Sca_{\ssl\ssr}^{\ssf\ssf}(s) = \Sca_{\ssr\ssl}^{\ssf\ssf}(s) &=& 
 \frac12 \, \Bigl(a_e^2 + b_e^2 \Bigr) \; s \, (t-u) \; F(s) ,
\end{eqnarray}
\begin{eqnarray}
\label{eeeeexpr2}
\Sca_{\ssl\ssl}^{\ssf\ssp}(s) = \Sca_{\ssl\ssl}^{\ssp\ssf}(s) 
= \Bigl( \Sca_{\ssl\ssr}^{\ssp\ssf}(s) \Bigr)^* =
\Bigl(\Sca_{\ssr\ssl}^{\ssf\ssp}(s)\Bigr)^*
&=& g_{e\ssl} \, (a_e +i b_e) \, \sqrt{stu} \; F(s) \nonumber\\
\Sca_{\ssl\ssr}^{\ssf\ssp}(s) = \Sca_{\ssr\ssl}^{\ssp\ssf}(s) 
= \Bigl( \Sca_{\ssr\ssr}^{\ssp\ssf}(s) \Bigr)^* =
\Bigl(\Sca_{\ssr\ssr}^{\ssf\ssp}(s)\Bigr)^*
&=& g_{e\ssr} \, (a_e +i b_e) \, \sqrt{stu} \; F(s) \nonumber\\
\end{eqnarray}
and
\begin{eqnarray}
\label{eeeeexpr3}
\Sca_{\ssl\ssl}^{\ssp\ssp}(s) &=& u \, \left\{ 
 \left[ {e^2 \over s} + \left({e\over \sw\cw}\right)^2 \; 
 {h_{e\ssl}^2 \over s - \mz^2} \right] 
 + \left[ g_{e\ssl}^2 
 +2 \kappa^2 \; (4u + 3s) \right] F(s) \right\} \nonumber\\
\Sca_{\ssr\ssr}^{\ssp\ssp}(s) &=& u \, \left\{
 \left[{e^2 \over s} + \left({e\over \sw\cw}\right)^2 
 \; {h_{e\ssr}^2 \over s - \mz^2} \right] 
 + \left[ g_{e\ssr}^2 
 + 2 \kappa^2 \; (4u + 3s) \right] F(s) \right\} \\
\Sca_{\ssl\ssr}^{\ssp\ssp}(s) = \Sca_{\ssr\ssl}^{\ssp\ssp}(s) &=& t \, \left\{
 \left[{e^2 \over s} + \left({e\over \sw\cw}\right)^2 \; 
 {h_{e\ssl}\, h_{e\ssr} \over s - \mz^2} \right] 
 + \left[ g_{e\ssl} \, g_{e\ssr} 
 + 2 \kappa^2 \; (4u + s) \right] F(s) \right\} \nonumber
\end{eqnarray} 
The $t$-channel amplitudes are obtained from these through the
replacement $(s,t,u) \to (t,s,u)$.

\subsection{Discussion}

The cross sections for graviphoton-exchange processes share many
features which are similar to those shown by the graviphoton
production cross sections. In particular, those interactions involving
the nonderivative (dimension-four) couplings can dominate both
graviphoton exchange through the magnetic-moment interaction and
graviton exchange. It can so dominate for two reasons:

\begin{enumerate}
\item Since the nonderivative interaction involves fewer derivatives,
  it is less suppressed by powers of $s/M_s^2$ than are the others.
\item Since it has the same helicity properties as have the
  Standard-Model photon and $Z$ couplings, it -- like the graviton but
  unlike the magnetic moment interaction -- can interfere with the
  Standard-Model amplitudes, and so contribute linearly in the cross
  section.
\end{enumerate}

Despite this relative enhancement, even tree-level dimension-four
graviphoton couplings cannot yet be ruled out by accelerator
experiments. This is because, even at the $Z$ peak, $Z$-graviphoton
interference gives corrections which are of order $M_\ssz^2/M_s^2
\lsim 1\%$ so long as $M_s \gsim 1$ TeV.

\section{Summary and Discussion}

There has been a great deal of interest in the possibility that extra
spatial dimensions could be quite large, and that the scale of quantum
gravity could be as low as about a TeV. Although supersymmetry is not
a prerequisite for such a scenario, it is nevertheless true that its
most realistic framework, M theory, does possess supersymmetry. In
this paper, we have examined the consequences of supersymmetry for
low-scale quantum gravity.

Since the graviton lives in a higher-dimensional space, it is
represented by a supermultiplet for a higher-dimensional
supersymmetry. In four dimensions, this supermultiplet includes
several particles, including (at least) one spin-one boson, the
graviphoton. Furthermore, because the graviphoton is related to the
graviton by supersymmetry, and since its couplings are gravitational
in strength, its mass is of order $M_s^2 / M_p$, where $M_s$ is the
scale of supersymmetry breaking and $M_p$ is the four-dimensional
Planck mass. Therefore, just as the graviton can be represented in
four dimensions as a tower of Kaluza-Klein states, so can the
graviphoton, with the principal difference being that, while the
lowest-mass graviton state has $m=0$, the lowest-mass graviphoton
state has $m = M_s^2 / M_p \sim 10^{-3}$ eV for $M_s \sim 1$ TeV. The
presence of the graviphoton can have important effects on the
experimental signatures for low-scale quantum gravity.

We must stress that, while various aspects of supersymmetry are
model-dependent (e.g.\ the number of supersymmetries, whether or not
they are broken on the brane, etc.), the existence of the graviphoton
is not. Thus, any supersymmetric model of TeV-scale quantum gravity
must take into account the effects due to the presence of the
graviphoton, in particular the constraints on graviphoton emission.

We consider two possibilities for the coupling of the graviphoton to
ordinary matter. First, the coupling can take the form of a
dimension-5 magnetic-moment term. In this case, purely on dimensonal
grounds, one expect rates for processes involving graviphotons to be
of the same order as those involving gravitons. After all, for a
graviphoton dimension-5 coupling to matter, both the graviton and
graviphoton are derivatively coupled, and the coupling constants for
both particles have the same dimensions.

The second, more interesting possibility is that the graviphoton can
couple to ordinary matter via dimension-4 vector and/or axial-vector
interactions. This can arise if ordinary matter has nonzero charges
under the $U(1)$ group gauged by graviphotons. More importantly, even
if those charges are zero, dimension-4 graviphoton couplings will be
generated, in general, by the mixing of the graviphoton with the
neutral standard-model gauge bosons. Since this kind of mixing does
not decouple, it will happen so long as {\it any} states carry both
graviphoton and ordinary Standard-Model charges, {\it regardless of
  how massive these states might be}. Therefore, we expect, on quite
general grounds, that the dimension-4 graviphoton couplings are at
least of loop size, and may be larger.

Once these dimension-4 couplings are generated, there are very
stringent constraints on the number of large extra dimensions and/or
the scale of quantum gravity, $M_s$. The point is the following. A key
feature of low-scale quantum gravity is the possibility of emitting
bulk-space particles (like gravitons and graviphotons) when ordinary
particles collide. This process permits ordinary matter to lose energy
into what are effectively invisible degrees of freedom.  But the
agreement between the measured rate of energy loss of the supernova SN
1987A with theoretical expectations due to neutrino emission provides
a bound on the rate of energy loss into other invisible species.  For
gravitons it was found that this constraint permits $M_s \sim 1$ TeV
only if the number of extra dimensions, $n$, satisfies $n > 2$
\cite{supernovae}.

If graviphotons are also present, then they will also contribute to
the energy loss of supernovae. However, all processes involving
gravitational particles are suppressed by powers of $E/M_s$, where $E$
is the energy scale of the process in question. (This is why it is
possible for low-scale gravity effects to have remained undetected
until now.) But should the graviphoton have a dimension-4 coupling to
matter, the cross section for its emission is enhanced by a factor
$(M_s/E)^2$ compared to that for graviton emission. Since the
temperature inside a supernova is around 30 MeV, this enchancement
factor is enormous: the rate for graviphoton emission can be larger
than that for graviton emission by about $10^{12}$. This leads to
considerably stronger supernova constraints on $n$ and/or $M_s$.

The strongest bounds arise if the dimension-4 couplings arise at tree
level (i.e.\ the ordinary fermions have charges under the graviphoton
$U(1)$). In this case we find that $M_s \sim 1$ TeV is permitted only
for $n \ge 5$. If one assumes that the dimension-4 couplings are
instead only of one-loop size, then the constraints are slightly
weaker: $n \ge 4$ is required in order to have $M_s \sim 1$ TeV. We
must stress that these are very conservative results. We have allowed
for an error of a factor of 10 in our supernova estimates, and we have
ignored the fact that, for larger values of $n$, the graviton
supermultiplet typically contains several graviphotons, and so can
have even stronger constraints.

We therefore conclude that, for models of low-scale quantum gravity
which include supersymmetry, either the number of large extra
dimensions is large, at least $n=4$, or there must be some symmetry
which prevents the generation of low-energy dimension-4 graviphoton
couplings to ordinary matter.

These constraints also affect the prospects for experimentally
detecting gravitational effects at accelerators. As described above,
one of the most important processes for detection is the emission of a
final-state gravitational particle. However, the cross section for the
process $f\ol{f} \to \gamma + GP$ is proportional to $(E/M_s)^n$.
Thus, for energies $E$ well below the gravity scale, larger values of
$n$ lead to smaller cross sections. (That is, paradoxically, the more
large extra dimensions there are, the harder it is to detect them).
Note that this holds both for graviphotons (with dimension-4 and/or
dimension-5 couplings) and gravitons.

This is shown explicitly in Figs.~(2) through (5): we see that larger
values of $n$ lead to considerably smaller cross sections. However, we
also note that, should a gravitational signal be found, one might hope
to distinguish between the spin-1 graviphoton and the spin-2 graviton
by the angular distribution of the emitted photon.

Finally, the graviphoton can also contribute virtually to
fermion-antifermion scattering. Unlike gravitational emission, this
contribution is independent of the number of large extra dimensions.
However, a word of caution is appropriate here: should a signal of
gravitational new physics be found in the measurement of $f{\bar f}
\to g {\bar g}$, it is unlikely that we will be able to identify the
type of virtual state responsible. This is because, in addition to the
graviton and graviphoton, higher string modes can also contribute to
the scattering at the same order in $1/M_s$ \cite{OpPollution}.

As usual, the contributions of the graviton and graviphoton are
suppressed by powers of $E/M_s$. The contributions of the graviphoton
with dimension-5 couplings to matter are quite similar in size to
those of the graviton. However, as above, the contributions of the
graviphoton with dimension-4 couplings are larger than those of the
graviton by a factor $M_s^2 / s$ in the amplitude. Even so, it is not
possible to detect the presence of a graviphoton in low-energy
experiments. Even at the $Z$ peak, taking into account $Z$-graviphoton
interference, gravitational correctons are at most of order
$M_Z^2/M_s^2 \sim 1\%$. Thus, if one hopes to detect virtual
gravitational contributions, one really needs experiments where the
centre-of-mass energy approaches the scale of quantum gravity.

Finally, we remark that one can learn something about the nature of
the virtual contributions to fermion-antifermion scattering if one can
polarize the initial fermions and measure the helicities of the
final-state fermions. This can be seen by examining the graviphoton
couplings. The dimension-4 couplings are helicity-preserving (as are
the graviton couplings), while the dimension-5 couplings are helicity
flipping. Thus a cross-section measurement as a function of helicity
is a powerful diagnostic of the presence of large extra dimensions.

\bigskip

\centerline{\bf Acknowledgements}
We thank Karim Benakli for helpful discussions. This research was
supported in part by U.S. DOE contracts DE-FG02-94ER40817 (ISU), NSERC
of Canada, and FCAR du Qu\'ebec.

\end{document}